\documentclass[physrev,superscriptaddress]{revtex4-2}   	
\usepackage{graphicx}				
\usepackage{amsmath}
\usepackage{natbib}
\usepackage{subcaption}
\usepackage{hyperref}
\hypersetup{
    colorlinks=true,
    linkcolor=blue,
    citecolor=blue,
    filecolor=magenta,      
    }
\usepackage{xcolor}
\usepackage{booktabs}

\begin{document}
\title{Electric interface condition for sliding and viscous contacts}
\author{J. Rekier}
\email[Corresponding author: ]{jeremy.rekier@observatory.be}
\affiliation{Royal Observatory of Belgium, 3 avenue circulaire, B-1180 Brussels, Belgium}
\affiliation{Department of Earth and Planetary Science, University of California, Berkeley, CA 94720, USA}
\author{S. A. Triana}
\affiliation{Royal Observatory of Belgium, 3 avenue circulaire, B-1180 Brussels, Belgium}
\author{A. Trinh}
\affiliation{Lunar and Planetary Lab, University of Arizona Tucson, AZ 85721-0092, USA}
\author{B. Buffett}
\affiliation{Department of Earth and Planetary Science, University of California, Berkeley, CA 94720, USA}
\date{\today}							
\begin{abstract}
First principles of electromagnetism impose that the tangential electric field must be continuous at the interface between two media. The definition of the electric field depends on the frame of reference leading to an ambiguity in the mathematical expression of the continuity condition when the two sides of the interface do not share the same rest frame. We briefly review the arguments supporting each choice of interface condition and illustrate how the most theoretically consistant choice leads to a paradox in induction experiments. We then present a model of sliding contact between two solids and between a fluid and a solid, and show how this paradox can be lifted by taking into account the shear induced by the differential motion in a thin intermediate viscous layer at the interface, thereby also lifting the ambiguity in the electric interface condition. We present some guidelines regarding the appropriate interface condition to employ in magnetohydrodynamics applications, in particular for numerical simulations where sliding contact is used as an approximation to the viscous interface between a conducting solid and a fluid of very low viscosity such as in planetary interior simulations. 
\end{abstract}
\maketitle

\section{Introduction}
Solving problems of electromagnetism in continuous media requires imposing continuity conditions at interfaces between regions of different properties. These interface conditions can be derived axiomatically from Maxwell's equations in the integral form \citep[e.g.][]{Jackson1999,Griffiths2013}. From the same starting point, one can derive Maxwell's equation in their differential \textit{local} form by making use of the divergence and curl theorems. Some textbooks prefer to work backward from there and attempt to derive interface conditions by integrating the differential Maxwell equations across the interface \citep[e.g.][]{Feynman1977}. While this is perfectly fine in most situations, this leads to difficulties when the two sides of the interface are not at rest with respect to each other. Such situation occurs in deforming media for which careful inspection of Maxwell's equations in their `material form' show that it is the tangential electric field measured in the instantaneous rest frame--equivalent to the \textit{electromotive force} per unit length--on each side of the interface that must be continuous \citep[][see also \citealp{Moffatt1978,EringenMaugin1990}]{LaxNelson1976}. This condition is relevant to electroelastic applications \citep{NelsonLax1976}. It is also the one commonly used in plasma physics \citep[][see also \citealp{ThorneBlandford2017}]{BernsteinEtAl1958,Lehnert1997,Schnack2009}. 

Recently, \citet{SatapathyHsieh2013} used two railgun experiments to confirm the validity of that condition at the interface between two solids in sliding contact. They showed that the commonly used alternative interface condition, which imposes that the tangential electric field measured \textit{in the lab frame}, be continuous failed to reproduce experimental data. This contrasts with the results of other classical experiments where the latter was found valid \citep[e.g.][]{HerzenbergLowes1957}. The present paper proposes to address this tension.

In Sec.~\ref{sec:theory} we state the problem in the mathematical form and briefly review the arguments in favour of both interface conditions. We also illustrate how the more theoretically consistent choice of boundary condition leads to a paradox in a simple rotating cylinder induction experiment. In Sec.~\ref{sec:solidinterface} we introduce a simple model of interface and show how the paradox is lifted when one takes into account the role of shears in an intermediate viscous layer between the two media. Finally, in Sec.~\ref{sec:solid-fluid} we extend the model to the case of a solid-fluid interface and show how the fluid boundary layer naturally serves the role of intermediate layer in that case. We present some conclusions and guidelines pertaining to magnetohydrodynamics applications and numerical simulations in Sec.~\ref{sec:discussion}.

\section{Theoretical motivation}
\label{sec:theory}

In this section, we review the electric boundary condition between two media from a theoretical point of view. We then illustrate its different interpretations using a simple example.

\subsection{Electric boundary condition}

At the interface between two media, the electric field, $\mathbf{E}$, satisfies the following junction condition \citep[e.g.][]{Jackson1999,Griffiths2013}:
\begin{equation}
\hat{\mathbf{n}}\times[\mathbf{E}]^+_-=\mathbf{0}~,
\label{eq:Econtinuity}
\end{equation}
where the notation $[\cdot]^+_-$ denotes the jump in the quantity in brackets across the interface with unit normal vector $\hat{\mathbf{n}}$. Equation \eqref{eq:Econtinuity} is ambiguous so long as one does not specify the frame of reference. Indeed, in the non-relativistic limit, the electric field, $\mathbf{E}'$, measured in the frame moving with velocity $\mathbf{v}$ is related to that measured in the lab frame, $\mathbf{E}$, via: $\mathbf{E}'=\mathbf{E}+\mathbf{v}\times\mathbf{B}$, where $\mathbf{B}$ is the magnetic field, which has the same value in both frames, i.e. $\mathbf{B}=\mathbf{B}'$. A careful inspection of Maxwell's equations in the integral form shows that Eq.~\eqref{eq:Econtinuity} should in fact read \citep[][see also \citealp{ThorneBlandford2017}]{Moffatt1978,EringenMaugin1990}:
\begin{align}
\hat{\mathbf{n}}&\times[\mathbf{E}']^+_-=\mathbf{0}\nonumber\\
\Leftrightarrow\hat{\mathbf{n}}&\times[\mathbf{E}+\mathbf{v}\times\mathbf{B}]^+_-=\mathbf{0}~.\label{eq:E+vxBcontinuity}
\end{align}
For convenience, we provide our own derivation of the above in Appendix \ref{sec:E+vxBcontinuity}. Equations \eqref{eq:Econtinuity} \& \eqref{eq:E+vxBcontinuity} are equivalent only when the two media are at rest relative to each other, i.e. when $[\mathbf{v}]^+_-=\mathbf{0}$. In particular, they are incompatible in situations where the two media are in \textit{sliding contact}. In such case, \citet{SatapathyHsieh2013} recently demonstrated the validity of Eq.~\eqref{eq:E+vxBcontinuity} based on two railgun experiments. Nonetheless, Eq. \eqref{eq:Econtinuity} seems to be favored in most applications. A case in point being the experiment of \citet{HerzenbergLowes1957} who studied induction in a rotating cylinder permeated by a uniform magnetic field. 

In the remainder of this section, we illustrate the difference between Eqs.~\eqref{eq:Econtinuity} and \eqref{eq:E+vxBcontinuity} using a simple model resembling the experiment of \citeauthor{HerzenbergLowes1957}. For this purpose, we will need the following additional condition which is always true:
\begin{equation}
\hat{\mathbf{n}}\cdot[\mathbf{B}]^+_-=0~.
\label{eq:normalBcontinuity}
\end{equation}
it will also be useful to rewrite Eqs.~\eqref{eq:Econtinuity} and \eqref{eq:E+vxBcontinuity} so as to avoid explicit mentions of the electric field. Using \textit{Ohm's law}, $\mathbf{j}=\sigma(\mathbf{E}+\mathbf{v}\times\mathbf{B})$, where $\mathbf{j}$ is the electric current density and $\sigma$ is the electric conductivity, combined with \textit{Amp\`ere's law} in the pre-Maxwell form relevant in magnetohydrodynamics applications \citep{Jackson1999}, $\mathbf{\nabla}\times\mathbf{B}=\mu\mathbf{j}$, where $\mu$ is the magnetic permeability, Eq.~\eqref{eq:Econtinuity} becomes:
\begin{equation}
\hat{\mathbf{n}}\times[\eta\mathbf{\nabla}\times\mathbf{B}]^+_-=(\hat{\mathbf{n}}\cdot\mathbf{B})[\mathbf{v}]^+_--[(\hat{\mathbf{n}}\cdot\mathbf{v})\mathbf{B}]^+_-~,
\label{eq:currentdiscontinuity}
\end{equation}
where we used Eq.~\eqref{eq:normalBcontinuity} and we introduced the magnetic diffusivity $\eta\equiv(\sigma\mu)^{-1}$. In this work, we focus on the case where $(\hat{\mathbf{n}}\cdot\mathbf{v})=0$, i.e. we take the boundary as static and impermeable to the velocity, allowing us to discard that last term of \eqref{eq:currentdiscontinuity}. By contrast, Eq.~\eqref{eq:E+vxBcontinuity} reduces to:
\begin{equation}
\hat{\mathbf{n}}\times[\eta\mathbf{\nabla}\times\mathbf{B}]^+_-=\mathbf{0}~.
\label{eq:currentcontinuity}
\end{equation}
Table~\ref{tab:C1C2} summarises the interface conditions used throughout this work.
\setlength\tabcolsep{0pt}
\begin{table}
  \centering
  \begin{tabular*}{0.78\linewidth}{@{\extracolsep{\fill}} cccc }
    \toprule
    Label & Condition & Equivalent form & Eqs. in main text\\
    \midrule
    C.I & $\hat{\mathbf{n}}\times[\mathbf{E}]^+_-=\mathbf{0}$ & $\hat{\mathbf{n}}\times[\eta\mathbf{\nabla}\times\mathbf{B}]^+_-=(\hat{\mathbf{n}}\cdot\mathbf{B})[\mathbf{v}]^+_-$ & Eqs.~\eqref{eq:Econtinuity}, and \eqref{eq:currentdiscontinuity} with $\hat{\mathbf{n}}\cdot\mathbf{v}=0$\\ 
    C.II & $\hat{\mathbf{n}}\times[\mathbf{E}+\mathbf{v}\times\mathbf{B}]^+_-=\mathbf{0}$ & $\hat{\mathbf{n}}\times[\eta\mathbf{\nabla}\times\mathbf{B}]^+_-=\mathbf{0}$ & Eqs.~\eqref{eq:E+vxBcontinuity} and~\eqref{eq:currentcontinuity} \\ 
    \bottomrule
  \end{tabular*}
  \caption{Interface conditions used throughout this work}
  \label{tab:C1C2}
\end{table}

Finally, we also make use of the following definition of the surface current density, $\mathbf{j}_s$, in terms of the tangential component of the magnetic field \citep{EringenMaugin1990}:
\begin{equation}
\hat{\mathbf{n}}\times[\mathbf{B}/\mu]^+_-=\mathbf{j}_s~.
\label{eq:surfacecurrentdef}
\end{equation}

\subsection{Paradox: induction in a solid cylinder}
\label{sec:cylinder}

\begin{figure}
\centering
\begin{subfigure}{0.32\textwidth}
\centering
\includegraphics[width=0.95\textwidth]{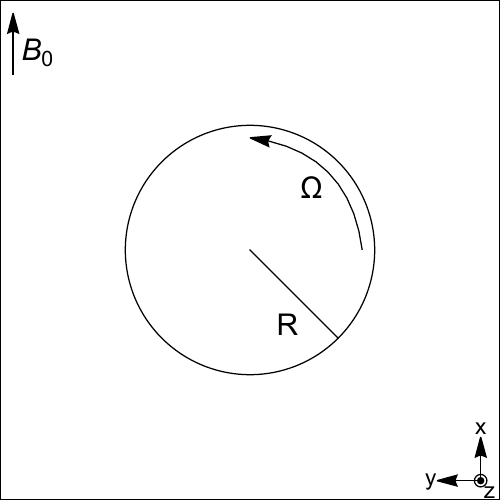}
\caption{Geometry \label{fig:cylinder_geometry}}
\end{subfigure}
\begin{subfigure}{0.32\textwidth}
\centering
\includegraphics[width=0.95\textwidth]{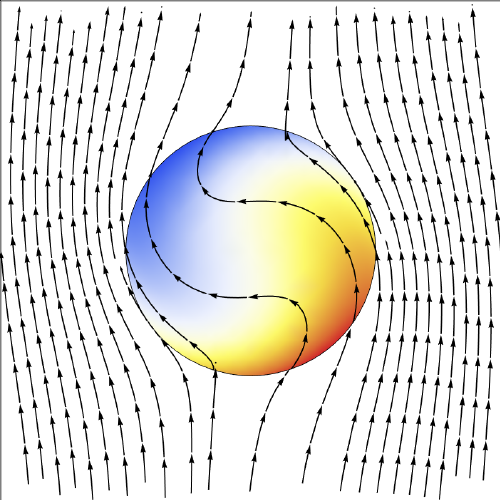}
\caption{(C.I) and $\hat{\mathbf{n}}\times[\mathbf{B}]^+_-=\mathbf{0}$ \label{fig:cylinder_Rm_10}}
\end{subfigure}
\begin{subfigure}{0.32\textwidth}
\centering
\includegraphics[width=0.95\textwidth]{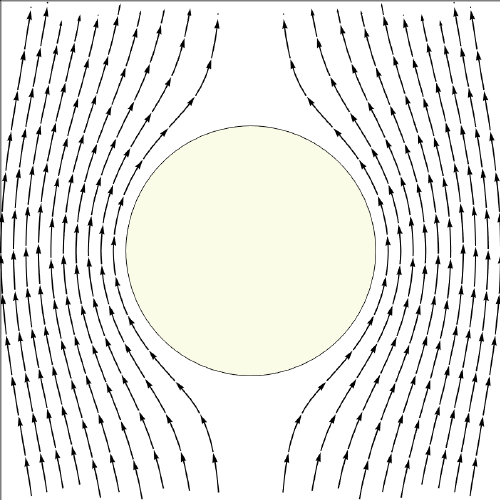}
\caption{(C.II) and $\hat{\mathbf{n}}\times[\mathbf{B}]^+_-\neq\mathbf{0}$\label{fig:cylinder_Rm_inf}}
\end{subfigure}
\caption{(\subref{fig:cylinder_geometry}) A rotating cylinder is immersed within a transverse uniform background magnetic field. (\subref{fig:cylinder_Rm_10}) Solution for $\mathrm{Rm}=10$, based on (C.I) and assuming continuity of the tangent magnetic field at the boundary. Streamlines show the magnetic field, colors show the normalized value of the axial current density with blue and red pointing respectively inside and outside the page. (\subref{fig:cylinder_Rm_inf}) Same thing but based on (C.II), and allowing for surface currents. The internal magnetic field is zero in that case for all values of $\mathrm{Rm}$, and the current density is restricted to a vanishingly thin sheet at the cylinder's surface.}
\label{fig:cylinders}
\end{figure}

We focus on the simple model introduced by \citet{Moffatt1978} and shown on Fig.~\ref{fig:cylinder_geometry} \citep[see also][]{Davidson2001}.
A cylindrical region filling the domain $s\in[0,R]$, with $s$ being the cylindrical radial coordinate, rotates steadily around the $z$-axis at the angular velocity $\Omega$. The whole volume is permeated by a transverse uniform external magnetic field, $\mathbf{B}_0$, along the $x$-direction. For simplicity, we assume that the cylinder and surrounding space have equal magnetic diffusivity, $\eta$. We solve the induction equation:
\begin{equation}
\partial_t\mathbf{B}=\mathbf{\nabla}\times(\mathbf{v}\times\mathbf{B})+\eta\nabla^2\mathbf{B}~,
\label{eq:induction}
\end{equation}
assuming $\partial_t\mathbf{B}=\mathbf{0}$ (steady regime). The system is then controlled by the value of the dimensionless \textit{magnetic Reynolds number} here defined as $\mathrm{Rm}\equiv\Omega R^2/\eta$. The solution for the magnetic field can be advantageously expressed as the real part of $\mathbf{B}=\mathbf{\nabla}\times(A_z\hat{\mathbf{z}})$, with \citep[see][for details]{Moffatt1978}:
\begin{equation}
A_z = B_0e^{i\varphi}
\begin{cases}
  c_1 J_1\left((1-i)\sqrt{\frac{\mathrm{Rm}}{2}}\frac{s}{R}\right), & s\leq R~, \\
  c_2 s^{-1}-s, & s>R~,
\end{cases}
\label{eq:Azsolcylinder}
\end{equation}
where $\varphi$ is the cylindrical azimutal coordinate, and $J_n$ denotes the Bessel function of degree $n$. The two integration constants, $c_1$ and $c_2$, are determined by boundary conditions. Equation \eqref{eq:normalBcontinuity} yields a single relation between $c_1$ and $c_2$, and a second one is needed. 

In the steady regime considered, $\mathbf{E}=\mathbf{0}$ both inside and outside of the cylinder, as can be checked readily from Eq.~\eqref{eq:Azsolcylinder}--combined with Amp\`ere's and Ohm's laws. Condition (C.I) therefore provides no constraint on the solution. In that same regime, it is then usual to consider that any surface currents at the interface have had ample time to diffuse into the volume such that, from Eq.~\eqref{eq:surfacecurrentdef}: $\hat{\mathbf{n}}\times[\mathbf{B}]^+_-=\mathbf{0}$. One then finds:
\begin{align}
c_1=-\frac{2 R}{\xi J_0(\xi )}~,&&c_2=R^2\left(1-\frac{2J_1(\xi)}{\xi J_0(\xi)}\right)~,\label{eq:solcylinder1}
\end{align}
where we have defined $\xi=(1-i)(\mathrm{Rm}/2)^{1/2}$. Figure~\ref{fig:cylinder_Rm_10} shows the corresponding solution for $\mathrm{Rm}=10$. By contrast, (C.II) \textit{does} provide a constraint on the solution and leads to:
\begin{align}
c_1=0~,&&c_2=R^2~.\label{eq:solcylinder2}
\end{align}
The corresponding solution is shown on Fig.~\ref{fig:cylinder_Rm_inf}. The most striking feature of this is the absence of magnetic induction inside the cylinder which behaves as a perfect conductor (as confirmed by taking the limit $\mathrm{Rm}\rightarrow\infty$ in Eq.~\ref{eq:solcylinder1}) even though we made no assumption on the value of $\mathrm{Rm}$ in arriving to Eq.~\eqref{eq:solcylinder2}. This solution also implies the existence of surface currents, as can be seen by injecting it back into Eq.~\eqref{eq:surfacecurrentdef}. A feature that seems unphysical in the steady regime for the same reason evoked above.


The inadequacy of (C.II) to reproduce the experimental results of \citeauthor{HerzenbergLowes1957} can be traced back to the assumption that the rotating cylinder and its surrounding are in sliding contact. In the experiment, the cylinder is maintained in electric contact with its supporting apparatus via a very thin layer of mercury which also serves as lubricant. Because of its finite viscosity, this intermediate layer becomes the host of intense shears during rotation. 
Section \ref{ref:transitionlayer} of the Appendix gives a formal derivation of the appropriate boundary condition in presence of such transition layer. In the following section we take a closer look at the role played by shears at the interface.

\section{The role of shear at the interface}
\label{sec:solidinterface}

\begin{figure}
\centering
\includegraphics[width=0.65\textwidth]{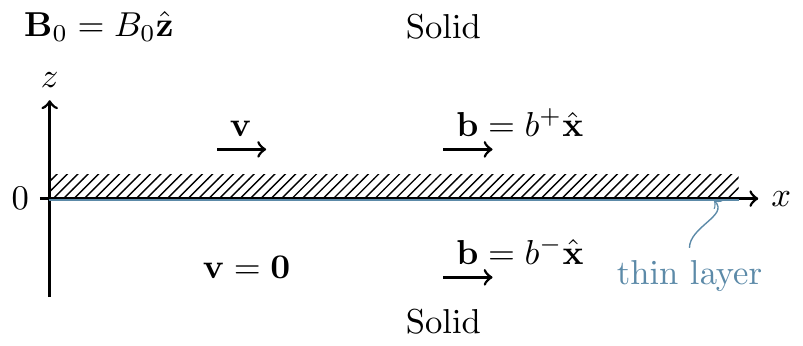}
\caption{Local model of the interface between two semi-infinite conducting solids sliding on top of each other. The presence of a thin viscous layer between the two solids (in blue) radically affects the solution (see main text).}
\label{fig:1D}
\end{figure}

We consider the simple toy-model depicted on Fig.~\ref{fig:1D} representing the interface between two conducting rigid media. 
The whole volume is permeated by a uniform external magnetic field, $\mathbf{B}_0$, pointing along the normal to the interface chosen as the $z$-axis, and the upper medium oscillates tangentially at a prescribed frequency, $\omega$, so that the velocity writes $\mathbf{v}=v\hat{\mathbf{x}}$, with:
\begin{equation}
v=
\begin{cases}
  \mathrm{Re}\left[\tilde{v}e^{-i\omega t}\right]~, &z\geq0 \\
  0~, &z<0
\end{cases}
~.\label{eq:velocity}
\end{equation}
For simplicity, we take the magnetic diffusivity as constant in the whole volume. We also focus exclusively on the region near the interface, so that we can safely approximate both sides as semi-infinite. The total magnetic field writes $\mathbf{B}=\mathbf{B}_0+\mathbf{b}$, where $\mathbf{b}$ is the perturbation induced by the oscillations which must obey the induction Eq.~\eqref{eq:induction}. We first look at the case where the thin viscous layer (in blue on the figure) is absent.


The induced field must point in the same direction as the velocity, and its magnitude can only depend on $z$, from symmetry. We thus write $\mathbf{b}=b\hat{\mathbf{x}}$, with:
\begin{equation}
b=
\begin{cases}
  \mathrm{Re}[\tilde{b}^+ e^{i(\lambda^+ z-\omega t)}], & z\geq0 \\
  \mathrm{Re}[\tilde{b}^- e^{i(\lambda^- z-\omega t)}], & z<0
\end{cases}
~.\label{eq:b+-}
\end{equation}
We obtain $\lambda^+$ and $\lambda^-$ by injecting the above into the induction Eq.~\eqref{eq:induction}, and imposing regularity of the solution at $z\rightarrow\pm\infty$, giving:
\begin{equation}
\lambda^\pm=\pm\frac{1+i}{\delta_*}~,
\end{equation}
where $\delta_*=\sqrt{2\eta/\omega}$ is the \textit{magnetic skin depth}. 

Finally, the values of the constants, $\tilde{b}^+$ and $\tilde{b}^-$, depend on our choice of boundary condition. As an attempt to make sense of the paradox of Sec.~\ref{sec:cylinder}, when (C.II) is imposed, we consider this condition first. In the present situation this gives the constraint:
\begin{equation}
\tilde{b}^+=-\tilde{b}^-~.
\end{equation}
Again, we see that the above strictly prohibits the continuity of the tangential magnetic field across the interface, except in the trivial case where $\tilde{b}^+=\tilde{b}^-=0$, and there is no induction between the two media. On the other hand, (C.I) can here be rewritten as:
\begin{equation}
[\eta\partial_zb+B_0v]^+_-=0~.
\label{eq:Econtinuity1D}
\end{equation}
This condition combined with the requirement of zero surface current, $\tilde{b}^+=\tilde{b}^-$, gives a non-zero value for the induced field on both sides:
\begin{equation}
\tilde{b}^\pm=(1+i)\frac{B_0\tilde{v}}{4\eta}\delta_*~.
\label{eq:sol1D}
\end{equation}
From the above, we see that (C.I) is clearly incompatible with (C.II) as they lead to very different solutions. Moreover it would seem that the former should be preferred over the latter on empirical ground, which is at odds with first principles, as well as plasma physics applications using (C.I). There is however a simple way to motivate the use of that condition physically if we consider the slightly modified situation where the thin blue intermediate viscous layer of Fig.~\ref{fig:1D} is present. We can model this by substituting the expression of the velocity Eq.~\eqref{eq:velocity} with one valid for the whole domain:
\begin{equation}
\mathbf{v}\rightarrow\Theta(z)\mathbf{v}
\end{equation}
where $\Theta(z)$ is the Heaviside distribution satisfying $\Theta(z\geq0)=1$, and $\Theta(z<0)=0$. Plugging into Eq.~\eqref{eq:induction}, keeping only linear terms in $\mathbf{v}$ and $\mathbf{b}$, and using the identity: $d\Theta/dz=\delta(z)$, gives:
\begin{equation}
\partial_t\mathbf{b}=B_0\delta(z)\mathbf{v}+\eta\nabla^2\mathbf{b}~,
\label{eq:inductiondelta}
\end{equation}
where $\delta(z)$ is the Dirac distribution. Integrating Eq.~\eqref{eq:inductiondelta} over an infinitesimal line segment across the interface \citep[see][]{Buffett1992} gives back Eq.~\eqref{eq:Econtinuity1D}, from which Eq.~\eqref{eq:sol1D} follows. This argument demonstrates the important role of viscous layers at the interface between two solids. It is valid so far as we can reasonably approximate that layer as infinitely thin but breaks down in situations where the conducting viscous fluid has a finite volume to which we now turn.

\section{Shear layer at a solid-fluid interface}
\label{sec:solid-fluid}

\begin{figure}
\centering
\includegraphics[width=0.65\textwidth]{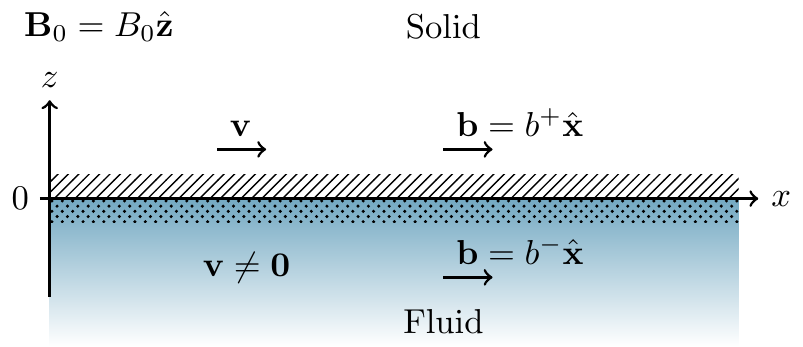}
\caption{Local model of solid-fluid interface. The finite viscosity of the fluid causes the appearance of thin boundary layer in the fluid region represented by the dotted area (see main text).}
\label{fig:1D-fluid}
\end{figure}

We consider the situation depicted on Fig.~\ref{fig:1D-fluid}. This is analogous to that of Sec.~\ref{sec:solidinterface}, except that the bottom region is now fluid and has finite viscosity. 
The flow velocity, $\mathbf{v}$, in the fluid region is governed by the momentum equation:
\begin{align}
\partial_t\mathbf{v}+(\mathbf{v}\cdot\mathbf{\nabla})\mathbf{v}&=-\frac{\nabla P}{\rho}+\frac{1}{\rho\mu}\left(\mathbf{\nabla}\times\mathbf{B}\right)\times\mathbf{B}+\nu\nabla^2\mathbf{v}~,\label{eq:NavierStokes}
\end{align}
where $\nu$ is the kinematic viscosity, $P$ is the pressure, and $\rho$ is the mass density taken as constant. $\mathbf{B}$ is the total magnetic field governed by the induction Eq.~\eqref{eq:induction}. We proceed as previously and write $\mathbf{B}=\mathbf{B}_0+\mathbf{b}$. The magnetic perturbation assumes the same form as Eq.~\eqref{eq:b+-}. The fluid velocity has a similar form, but is limited to the fluid region, so that the whole velocity field $\mathbf{v}=v\hat{\mathbf{x}}$ can be summarised by:
\begin{equation}
v=
\begin{cases}
  \mathrm{Re}\left[\tilde{v}^+e^{-i\omega t}\right], &z\geq0 \\
  \mathrm{Re}\left[\tilde{v}^- e^{i(\lambda^- z-\omega t)}\right]~, &z<0
\end{cases}
~.
\end{equation}
To linear order in $\mathbf{v}$ and $\mathbf{b}$, Eqs.~\eqref{eq:induction} and \eqref{eq:NavierStokes} may then be written in the following matrix form:
\begin{equation}
\left(
\begin{array}{cc}
v_A\lambda\sqrt{\rho\mu} & \omega+i\lambda^2\eta\\
\omega+i\lambda^2\nu & v_A\lambda/\sqrt{\rho\mu}
\end{array}
\right)
\left(
\begin{array}{c}
\tilde{b}\\
\tilde{v}
\end{array}
\right)=0~.
\label{eq:matrix}
\end{equation}
where we have dropped the superscripts from $\tilde{v}^-$, $\tilde{b}^-$, and $\lambda^-$, for readability, and we have introduced $v_A=B_0/\sqrt{\rho\mu}$, representing the \textit{Alfv\'en velocity} inside the fluid \citep{Alfven1942a}. The determinant of the matrix in Eq.~\eqref{eq:matrix} must be equal to zero. Solving for $\lambda$, we find \citep[see also][]{SchaefferEtAl2012}:
\begin{equation}
\lambda^2=-\left(\frac{v_A^2}{2 \eta  \nu }-i\omega
   \frac{\eta + \nu}{2 \eta  \nu}\right) \pm\sqrt{\frac{\omega^2}{\eta\nu}-\left(\omega  \frac{(\eta +\nu )}{2\eta\nu}+i \frac{v_A^2}{2\eta\nu}\right)^2}
\end{equation}
The numerical values of these numbers depend on the parameters. With geophysical applications in mind, we take values that are typical for the  Earth's core: $\rho=10^{4}~\text{kg}/\text{m}^3$, $\mu=4\pi\times10^{-7}~\text{H}/\text{m}$, $\eta=0.5~\text{m}^2/\text{s}$, $\nu=10^{-6}~\text{m}^2/\text{s}$, and $v_A=4.5\times10^{-3}~\text{m}/\text{s}$. 

\begin{figure}
\centering
\includegraphics[width=0.55\textwidth]{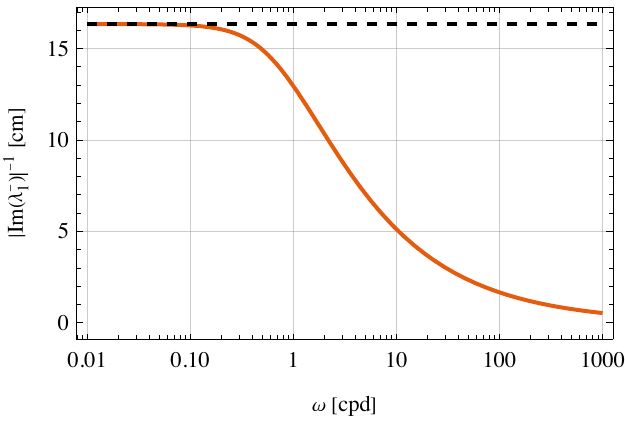}
\caption{Thickness of the boundary layer at the fluid-solid interface for parameters typical of the Earth's upper fluid core as a function of the frequency (in cycles per day). The dashed line shows the steady limit corresponding to the Hartmann layer thickness $\delta_H=\sqrt{\nu\eta}/v_A$.}
\label{fig:Hartmann-layer}
\end{figure}

In total, there are four solutions for $\lambda$ which we parametrize as (reintroducing the superscript):
\begin{equation}
\lambda^-=k-i\delta^{-1}~,
\end{equation}
where $k$ and $\delta$ are real numbers representing the wave number, and the decay length, respectively. We only keep solutions satisfying $\delta>0$, which ensures that the perturbations converge to zero at $z\rightarrow-\infty$. Among the two remaining solutions, there is one for which $k$ is large and $\delta$ is small representing a rapidly decaying short wavelength plane wave. For the other, $k$ is small and $\delta$ is large representing a slowly decaying plane wave with long wavelength:
\begin{subequations}
\begin{align}
\text{boundary layer:}&&\{\tilde{v}^-_1,\tilde{b}^-_1\},&&\lambda^-_1=k_1-i\delta^{-1}_1~,\label{eq:lambda1}\\
\text{travelling wave:}&&\{\tilde{v}^-_2,\tilde{b}^-_2\},&&\lambda^-_2=k_2-i\delta^{-1}_2~,\label{eq:lambda2}
\end{align}
\end{subequations}
with $\delta_1\ll\delta_2$ and $k_1\gg k_2$. In the first solution, $\delta_1$ corresponds to the thickness of the boundary layer illustrated by the dotted area on Fig.~\ref{fig:1D-fluid}. Its thickness is shown on Fig.~\ref{fig:Hartmann-layer} as a function of the frequency given in cycles per day (cpd). It is maximum in the steady limit where it is equal to the \textit{Hartmann layer} thickness $\delta_H=\sqrt{\nu\eta}/v_A$ (dashed line).
Like for the solid interface, the constants $\tilde{v}^-_1$, $\tilde{b}^-_1$, $\tilde{v}^-_2$, and $\tilde{b}^-_2$, as well as $\tilde{b}^+$ for the upper solid region, are set by the boundary conditions. The values of $\tilde{b}^-_1$ and $\tilde{b}^-_2$ are related to $\tilde{v}^-_1$ and $\tilde{v}^-_2$ by Eq.~\eqref{eq:matrix}, so that only three constraints are needed. One is given by the requirement that the velocity be continuous across the interface (no-slip). This renders the condition on the tangential electric field unambiguous, the latter serving as our second constraint. Finally, we assume that the tangential magnetic field is continuous at the interface (no surface currents). In our simple model, these three constraints write:
\begin{subequations}
\begin{align}
&\text{No-slip:}&\tilde{v}^-_1+\tilde{v}^-_2&=\tilde{v}^+~,\label{eq:noslip1D}\\
&\text{Current continuity, (C.I) and (C.II) equivalent:}&\lambda^-_1\tilde{b}^-_1+\lambda^-_2\tilde{b}^-_2&=\lambda^+\tilde{b}^+~,\label{eq:currentcontinuity1D}\\
&\text{Magnetic field continuity:}&\tilde{b}^-_1+\tilde{b}^-_2&=\tilde{b}^+~.
\end{align}
\end{subequations}
The solution is then completely determined once we set the value of $\tilde{v}^+$. For simplicity, we take $\tilde{v}^+=v_A$, which matches well the relative velocity between the Earth's core and mantle induced by the Earth's precession \citep{Tilgner2015}.

\begin{figure}
\centering
\begin{subfigure}[b]{0.49\textwidth}
\centering
\includegraphics[width=\textwidth]{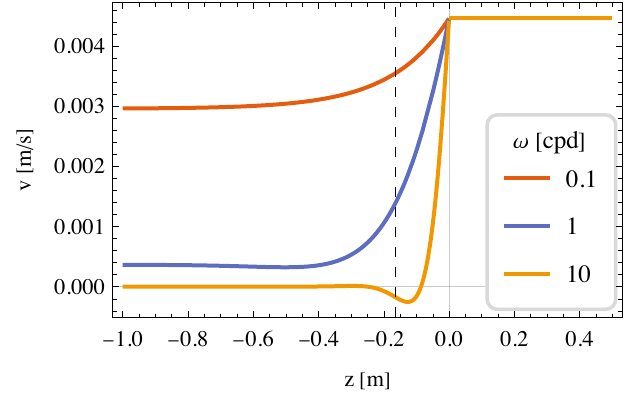}
\caption{Velocity}    
\label{fig:u}
\end{subfigure}
\hfill
\begin{subfigure}[b]{0.49\textwidth}
\centering
\includegraphics[width=\textwidth]{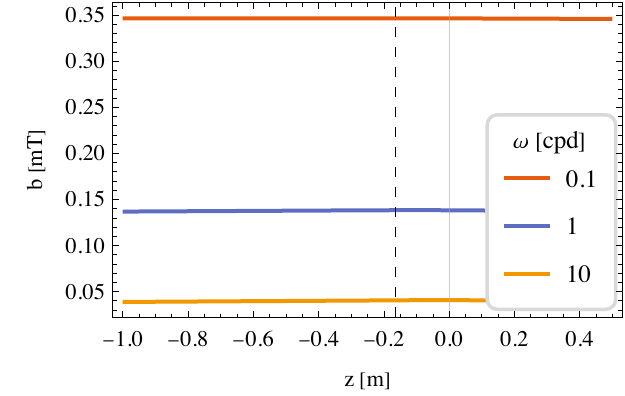}
\caption{Magnetic field}    
\label{fig:b}
\end{subfigure}
\vskip\baselineskip
\begin{subfigure}[b]{0.49\textwidth}   
\centering
\includegraphics[width=\textwidth]{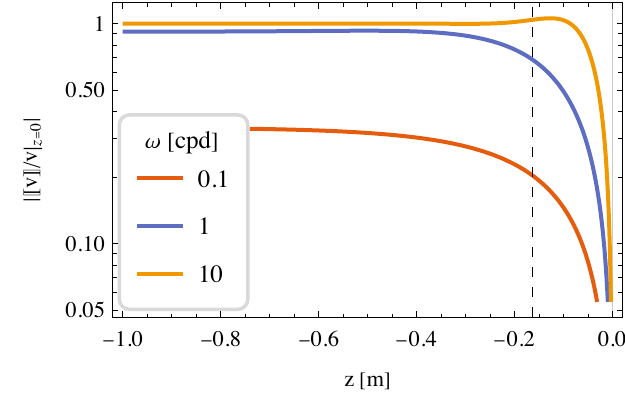}
\caption{Jump in velocity}
\label{fig:u_rel}
\end{subfigure}
\hfill
\begin{subfigure}[b]{0.49\textwidth}   
\centering
\includegraphics[width=\textwidth]{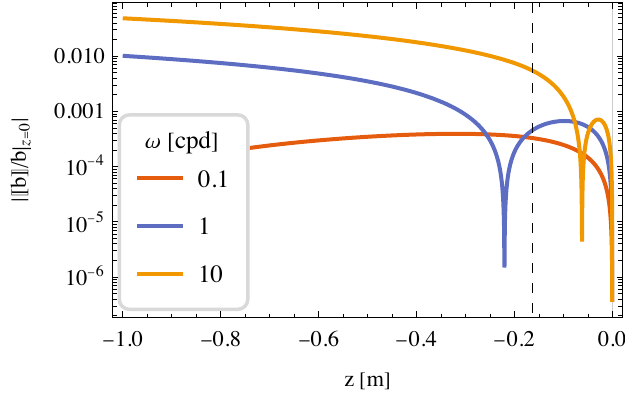}
\caption{Jump in magnetic field}
\label{fig:b_rel}
\end{subfigure}
\caption{Top row: velocity and magnetic field perturbations as a function of depth, for three values of the frequency (in cycles per day). Bottom row: relative jump in velocity and magnetic field. The black dashed lines indicate the depth of the Hartmann layer in the steady limit (see Fig.~\ref{fig:Hartmann-layer}).}
\label{fig:solution}
\end{figure}

The top row of Fig.~\ref{fig:solution} shows the magnitude of the velocity and magnetic field as a function of depth for three values of the frequency. 
The vertical dashed line on each plot indicates the width of the Hartmann layer in the steady limit (see Fig.~\ref{fig:Hartmann-layer}). While the effect of this boundary layer is clearly visible on the velocity profile, it is hardly discernable in the magnetic field perturbation. This is the expected behaviour for fluids where $\eta\gg\nu$ such as in the Earth's core \citep{Stewartson1960,Stewartson1960a}. In this case, the velocity of the flow just below the boundary layer converges quickly to its value in the bulk of the fluid---the so-called \textit{free stream}---which is here simply $v=0$. In order to better quantify changes across the boundary layer, we introduce the following notation \citep[see also][]{RobertsScott1965,AchesonHide1973}:
\begin{equation}
[[b]]\equiv b|_{z=0}-b~,
\end{equation}
and likewise for $v$, and other quantities. The jumps in $v$ and $b$ are shown on the bottom row of Figs.~\ref{fig:solution} as a function of depth, in absolute value and normalised by the values of those fields at the interface. 

\begin{figure}
\centering
\begin{subfigure}[b]{0.49\textwidth}
\centering
\includegraphics[width=\textwidth]{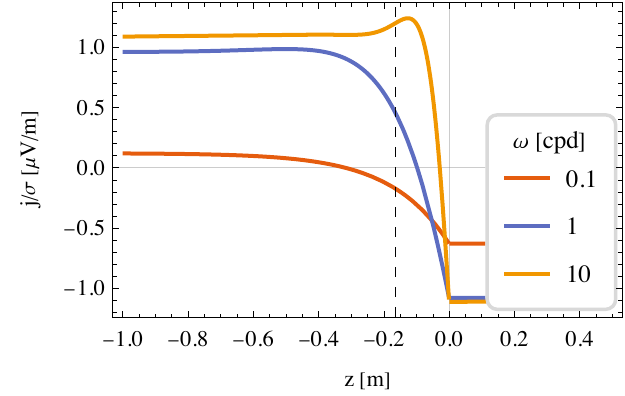}
\caption{Current density}    
\label{fig:j}
\end{subfigure}
\hfill
\begin{subfigure}[b]{0.49\textwidth}
\centering
\includegraphics[width=\textwidth]{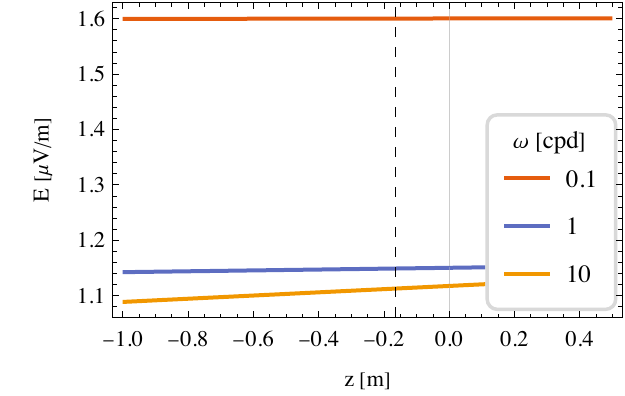}
\caption{Electric field}    
\label{fig:E}
\end{subfigure}
\vskip\baselineskip
\begin{subfigure}[b]{0.49\textwidth}   
\centering
\includegraphics[width=\textwidth]{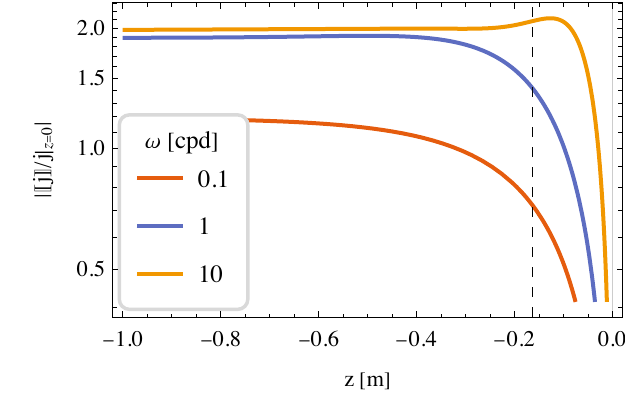}
\caption{Jump in current density}
\label{fig:j_rel}
\end{subfigure}
\hfill
\begin{subfigure}[b]{0.49\textwidth}   
\centering
\includegraphics[width=\textwidth]{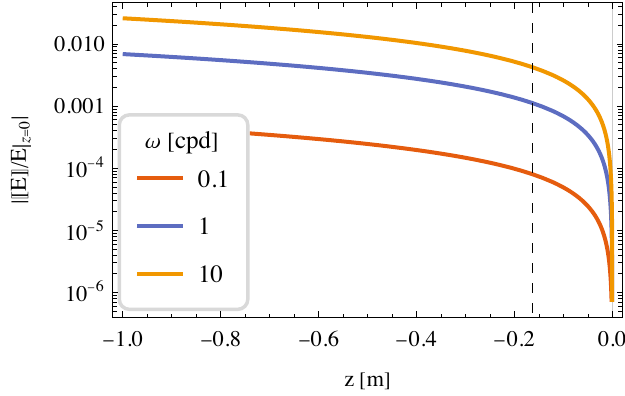}
\caption{Jump in electric field}
\label{fig:E_rel}
\end{subfigure}
\caption{Top row: electric current density divided by electric conductivity, and electric field near the interface. Bottom row: relative jump in current density and electric field. The black dashed lines indicate the depth of the Hartmann layer in the steady limit (see Fig.~\ref{fig:Hartmann-layer}).}
\label{fig:solutionjE}
\end{figure}

Finally, we can use our simple toy-model to probe which of (C.I) or (C.II) better represents the behavior of the electric field near the interface. The top row of Fig.~\ref{fig:solutionjE} shows the magnitude of the electric current density divided by the electric conductivity, $\mathbf{j}/\sigma=\mathbf{E}+\mathbf{v}\times\mathbf{B}_0$, and of the electric field, $\mathbf{E}$, as a function of depth for three values of the frequency. The bottom row shows the jump in these quantities divided by their values at the interface. 
We can see that the current density varies a lot across the boundary layer, whereas the electric field remains almost constant. We can express this mathematically as:
\begin{equation}
[[\eta\partial_zb+B_0v]]\approx0~,\label{eq:Econtinuity1D_2}
\end{equation}
which is valid within the boundary layer. Equation~\eqref{eq:Econtinuity1D_2} is analogous to Eq.~\eqref{eq:Econtinuity1D} derived in Sec.~\ref{sec:solidinterface} for the viscous solid-solid interface which we found to be equivalent (C.I). In this light, the latter can perceived as the limit case of Eq.~\eqref{eq:Econtinuity1D_2} when the boundary layer is infinitely thin. This result proves the validity of (C.I) at a solid-fluid interface when it is understood as a condition relating the values of the fields inside the solid layer to their values at the top of the free-stream, i.e. just below the thin boundary layer. 

\begin{figure}
\centering
\begin{subfigure}[b]{0.49\textwidth}
\centering
\includegraphics[width=\textwidth]{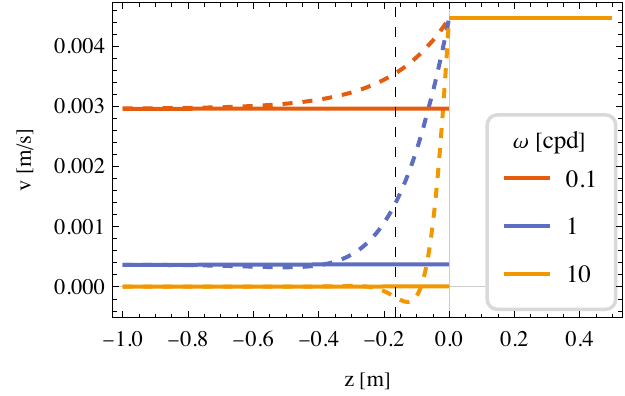}
\caption{Velocity}    
\label{fig:u_sf}
\end{subfigure}
\hfill
\begin{subfigure}[b]{0.49\textwidth}
\centering
\includegraphics[width=\textwidth]{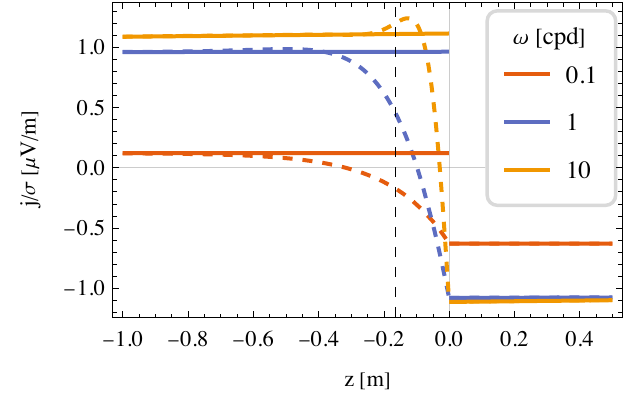}
\caption{Current density}    
\label{fig:j_sf}
\end{subfigure}
\caption{Velocity and electric current density near a solid-fluid interface using the boundary conditions Eqs.~\eqref{eq:stressfree1Dlambda} and \eqref{eq:Econtinuity1Dlambda} (plain curves) compared to the exact solution of Sec.~\ref{sec:solid-fluid} (dashed curves).}
\label{fig:solution_sf}
\end{figure}

As we have argued in Sec.~\ref{sec:solidinterface}, condition (C.I) automatically accounts for viscous shears at the interface and their effect on the fields. We can check that this is also true for the solid-fluid interface by slightly altering the above model, replacing the continuity (no-slip) condition on the velocity field at the interface with the a condition on its gradient: $[\partial_zv]^+_-=0$, which amounts to impose that the shear stresses are continuous across the interface. As the upper medium is here assumed perfectly rigid, the viscous stresses in the fluid must vanish at the boundary--- this is the so-called \textit{stress-free} condition. More specifically, this amounts to replace Eq.~\eqref{eq:noslip1D} and \eqref{eq:currentcontinuity1D} by:
\begin{subequations}
\begin{align}
&\text{Stress-free:}&\lambda_1^-\tilde{v}_1^-+\lambda_2^-\tilde{v}_2^-&=0~,\label{eq:stressfree1Dlambda}\\
&\text{Electric field continuity (C.I):}&\eta(\lambda_1^-\tilde{b}_1^-+\lambda_2^-\tilde{b}_2^--\lambda^+\tilde{b}^+)&=iB_0(v_1^-+v_2^--v^+)~,\label{eq:Econtinuity1Dlambda}
\end{align}
with Eq.~\eqref{eq:Econtinuity1Dlambda} reducing to Eq.~\eqref{eq:currentcontinuity1D}, only in the no-slip hypothesis.
\end{subequations}
Figure~\ref{fig:solution_sf} shows the solution for $u$, and the magnitude of $\mathbf{j}/\sigma$ using the above. The dashed curves correspond to the exact solution of Sec.~\ref{sec:solid-fluid} given for comparison.
We see that, even though shears have been removed from the model, the exact solution rapidly becomes equivalent to the approximate solution outside of the boundary layer by virtue of (C.I) alone. The approximate magnetic and electric fields are undistinguishable from the exact profiles shown on Figs.~\ref{fig:b} and \ref{fig:E} and are therefore not repeated here. 

\section{Conclusion}
\label{sec:discussion}

The continuity condition on the tangential part of the electric field at the interface between two media depends on the details of the problem. In situations where the two sides are in true sliding contact, theoretical as well as experimental considerations indicate that (C.II) should be preferred over (C.I). We have shown how the latter becomes relevant whenever an intermediary viscous layer is present between the two media. For a solid-fluid interface, this layer corresponds to the fluid boundary layer. Therefore, both geophysical and plasma physics applications (just to name a few) are equally justified in their respective usage of (C.I) and (C.II), with the former operating under the--often implicit--assumption that there are boundary layers present.

In practical fluid dynamics computations, the very small thickness of such layers can be challenging to model. This is certainly true when dealing with the Earth's fluid core in which case it is sometimes useful to get rid of the boundary layer entirely by assuming that the free stream extends all the way to the interface, thereby approximating the sharp jump in the flow velocity at the interface as a true discontinuity. In such case, we have shown in Sec.~\ref{sec:solid-fluid} that using (C.I) guarantees that the approximate solution will be close to the physical solution if the boundary layer of the latter is sufficiently thin (Fig.~\ref{fig:solution_sf}), even though the use of (C.I) is not strictly self-consistent from a theoretical point of view.

In summary, in order to decide which condition to use in a given application, one should start by assessing the presence and thickness of viscous layers at the interface considered. This assessment and its consequences for practical studies of the magnetohydrodynamics of the Earth's fluid core will be presented a future work.

\section*{Acknowledgement}
J. R. and S.T. express their warm gratitude to Veronique Dehant and Tim Van Hoolst for their encouragement and support, as well as Felix Gerick for enriching discussions. They also acknowledge financial support from the European Research Council (ERC) under the European Union’s Horizon 2020 research and innovation program (Synergy Grant agreement no. 855677 GRACEFUL). J.R. would like to thank UC Berkeley’s Department of Earth and Planetary Science for their hospitality during the making of this work. We also would like to thank the anonymous reviewer for their comments which helped to improve some of the arguments presented in the original manuscript.

\newpage 

\begin{appendix}
\numberwithin{equation}{section}

\section{Derivation of the electric boundary condition}
\label{sec:E+vxBcontinuity}

\subsection{Faraday's law of induction}
Faraday's law of induction relates the rate of change of the magnetic field flux through the (open) surface, $\mathcal{S}$, to the \textit{electromotive force}, acting on that surface, the latter being defined as the integral of the \textit{Lorentz Force} per unit charge over the enclosing contour, $\mathcal{C}$:
\begin{equation}
-\frac{d}{dt}\int_{\mathcal{S}(t)}\mathbf{B}\cdot\hat{\boldsymbol{n}}dS=\oint_{\mathcal{C}(t)}\left(\mathbf{E}+\mathbf{v}\times\mathbf{B}\right)\cdot\hat{\boldsymbol{t}}dl~,
\label{eq:Faraday}
\end{equation}
where $\hat{\boldsymbol{n}}$ and $\hat{\boldsymbol{t}}$ denote the unit vectors normal to $\mathcal{S}$ and tangent to $\mathcal{C}$, respectively. In general, these two domains may be time dependent, as reflected in the above notation. This prevents simply swapping the order of the integral and time derivative in Eq.~\eqref{eq:Faraday}. Instead we have, based on \textit{Reynolds's transport theorem} \citep[see \textit{e.g.}][]{Aris1990}:
\begin{equation}
-\frac{d}{dt}\int_{\mathcal{S}(t)}\mathbf{B}\cdot\hat{\boldsymbol{n}}dS=-\int_{\mathcal{S}(t)}\left(\frac{\partial\mathbf{B}}{\partial t}+\left(\mathbf{\nabla}\cdot\mathbf{B}\right)\mathbf{v}\right)\cdot\hat{\boldsymbol{n}}dS+\oint_{\mathcal{C}(t)}\left(\mathbf{v}\times\mathbf{B}\right)\cdot\hat{\boldsymbol{t}}dl~.
\label{eq:ReynoldsB}
\end{equation}
Inserting back into Eq.~\eqref{eq:Faraday}, and using $\mathbf{\nabla}\cdot\mathbf{B}=0$, we find:
\begin{equation}
-\int_{\mathcal{S}(t)}\frac{\partial\mathbf{B}}{\partial t}\cdot\hat{\boldsymbol{n}}dS=\oint_{\mathcal{C}(t)}\mathbf{E}\cdot\hat{\boldsymbol{t}}dl~.
\label{eq:Faradaymoving2}
\end{equation}
At this point, we must emphasise that Eq.~\eqref{eq:ReynoldsB} is only valid \textit{if both $\mathbf{B}$ and $\mathbf{v}$ are differentiable functions throughout the surface, $\mathcal{S}$}. If, in addition, the electric field, $\mathbf{E}$, is \textit{continuous} throughout the same surface, we may use \textit{Stokes's theorem} which in the present context reads:
\begin{equation}
\oint_{\mathcal{C}(t)}\mathbf{E}\cdot\hat{\boldsymbol{t}}dl=\int_{\mathcal{S}(t)}\mathbf{\nabla}\times\mathbf{E}\cdot\hat{\boldsymbol{n}}dS~.
\label{eq:Stokesthm}
\end{equation}
Combining Eqs.~\eqref{eq:Faradaymoving2} and \eqref{eq:Stokesthm}, we finally arrive to:
\begin{equation}
\int_{\mathcal{S}(t)}\left(\frac{\partial\mathbf{B}}{\partial t}+\mathbf{\nabla}\times\mathbf{E}\right)\cdot\hat{\boldsymbol{n}}dS=\mathbf{0}~.
\end{equation}
The above must be valid for any surface, $\mathcal{S}(t)$, which implies that:
\begin{equation}
\mathbf{\nabla}\times\mathbf{E}=-\frac{\partial\mathbf{B}}{\partial t}~,\label{eq:Maxwell}
\end{equation}
which is one of \textit{Maxwell's equations} in the local (differential) form. Crucially, the above derivation shows that Eq.~\eqref{eq:Maxwell} is independent of the reference frame. It also explains why it would be inconsistent to try to replace $\mathbf{E}$ by $\mathbf{E}+\mathbf{v}\times\mathbf{B}$ in the above equation. Doing so would miss the effect induced by the motion of the surface element that is implicit in Eq.~\eqref{eq:Maxwell} but explicit in Eq.~\eqref{eq:Faraday}. This makes the latter generally more suited to derive boundary conditions.

\subsection{Electric boundary condition}

\begin{figure}
\centering
\begin{subfigure}{0.49\textwidth}
\centering
\includegraphics[width=0.9\textwidth]{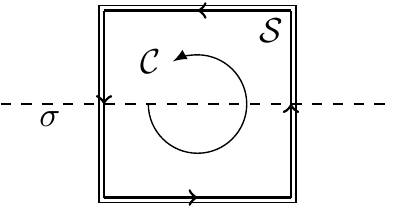}
\caption{\label{fig:JC1}}
\end{subfigure}
\begin{subfigure}{0.49\textwidth}
\centering
\includegraphics[width=0.9\textwidth]{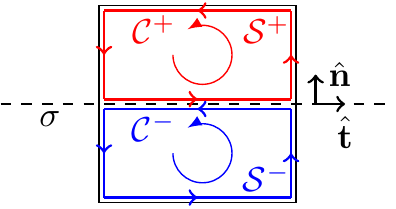}
\caption{\label{fig:JC2}}
\end{subfigure}
\caption{(\subref{fig:JC1}) the circuit contour, $\mathcal{C}$, enclosing the surface, $\mathcal{S}$, intersects the surface of discontinuity between two regions of space perpendicularly along the straight line, $\sigma$. (\subref{fig:JC2}) it is always possible to decompose each of $\mathcal{C}$ and $\mathcal{S}$ into two parts if we specify appropriate junction conditions at $\sigma$ (see main text).}
\label{fig:faradayloop1}
\end{figure}

In the previous section, we have assumed that the three vectors $\mathbf{v}$, $\mathbf{B}$, and $\mathbf{E}$ are continuous throughout the whole domain. We now look at what happens when this is not the case by considering the situation of Fig.~\ref{fig:JC1} where the surface, $\mathcal{S}$, intersects the interface between two regions of space perpendicularly along the dashed line, $\sigma$ with unit normal and tangent vectors, $\hat{\mathbf{n}}$ and $\hat{\mathbf{t}}$, which should not be confused with the previously introduced `dummy' vector variables, $\hat{\boldsymbol{n}}$ and $\hat{\boldsymbol{t}}$.
If $\mathbf{v}$, $\mathbf{B}$, and $\mathbf{E}$ are \textit{piecewise} continuous on both sides of the surface, we can do as on Fig.~\ref{fig:JC2} and divide the surface $\mathcal{S}$ into two parts, $\mathcal{S}^+$ and $\mathcal{S}^-$, such that $\mathcal{S}=\mathcal{S}^+\cup\mathcal{S}^-$ and we have the following property:
\begin{equation}
\int_\mathcal{S}\mathbf{A}\cdot\hat{\boldsymbol{n}}dS=\int_{\mathcal{S}^+}\mathbf{A}^+\cdot\hat{\boldsymbol{n}}dS+\int_{\mathcal{S}^-}\mathbf{A}^+\cdot\hat{\boldsymbol{n}}dS~,
\label{eq:S=S+S}
\end{equation}
which is true for any vector, $\mathbf{A}$, that is piecewise continuous on $\mathcal{S}^+$ and $\mathcal{S}^-$ where its value is denoted formally as $\mathbf{A}^+$ and $\mathbf{A}^-$, respectively. We have to be a little more careful in combining contour integrals \citep{Eringen1980}:
\begin{equation}
\oint_\mathcal{C}\mathbf{A}\cdot\hat{\boldsymbol{t}}dl=\oint_{\mathcal{C}^+}\mathbf{A}^+\cdot\hat{\boldsymbol{t}}dl+\oint_{\mathcal{C}^-}\mathbf{A}^-\cdot\hat{\boldsymbol{t}}dl-\int_\sigma[\mathbf{A}]^+_-\cdot\hat{\mathbf{t}} dl~,
\label{eq:C=C+C}
\end{equation}
where we have replaced $\hat{\boldsymbol{t}}$ by its constant value, $\hat{\mathbf{t}}$, in the last term as shown on Fig.~\ref{fig:JC2}, and we have used the following notation introduced in the main text:
\begin{equation}
[\mathbf{A}]^+_-\equiv(\mathbf{A}^+-\mathbf{A}^-)|_\sigma~.
\label{eq:brackets+-}
\end{equation}
In the special case where $\mathbf{A}$ is continuous throughout the whole domain, the two values, $\mathbf{A}^+$ and $\mathbf{A}^-$, converge to a single limit at $\sigma$ and the last term in Eq.~\eqref{eq:C=C+C} is zero.

Inserting Eqs.~\eqref{eq:S=S+S} and \eqref{eq:C=C+C} into Faraday's law Eq.~\eqref{eq:Faraday}, we may then repeat the exercise of the previous section to reduce the contour integrals on both sides of the interface into surface integrals which may then be combined into a single integral using Eq.~\eqref{eq:S=S+S}. The final result is:
\begin{equation}
\int_{\mathcal{S}(t)}\left(\frac{\partial\mathbf{B}}{\partial t}+\mathbf{\nabla}\times\mathbf{E}\right)\cdot\hat{\boldsymbol{n}}dS=\int_\sigma[\mathbf{E}+\mathbf{v}\times\mathbf{B}]^+_-\cdot\hat{\mathbf{t}} dl~.
\label{eq:Faradaysigma}
\end{equation}
Eq.~\eqref{eq:Faradaysigma} must be valid regardless of our choice of surface, $\mathcal{S}$. In particular, taking the limit $\mathcal{S}\rightarrow0$, we must have:
\begin{equation}
[\mathbf{E}+\mathbf{v}\times\mathbf{B}]^+_-\cdot\hat{\mathbf{t}}=0~,
\label{eq:EJCt}
\end{equation}
which is equivalent to (C.II).

\subsection{Electric boundary condition with a transition layer}
\label{ref:transitionlayer}

\begin{figure}
\centering
\includegraphics[width=0.45\textwidth]{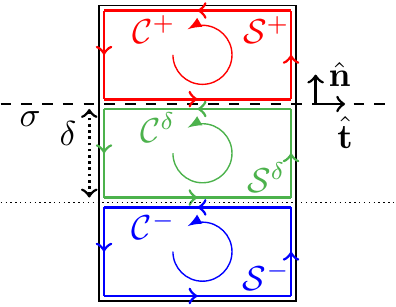}
\caption{Similar to Fig.~\ref{fig:JC2} but with an intermediate transition layer of thickness, $\delta$, with cross section, $\mathcal{S}^\delta$, enclosed by the contour, $\mathcal{C}^\delta$.}
\label{fig:faradayloop2}
\end{figure}

We now turn to the situation represented on Fig.~\ref{fig:faradayloop2} where, in addition to the two regions $\mathcal{S}^+$ and $\mathcal{S}^-$, there is an intermediate region of width, $\delta$, with cross-section, $\mathcal{S}^\delta$, enclosed by the contour, $\mathcal{C}^\delta$, and within which the values of all fields transition smoothly between those of $\mathcal{S}^+$ and $\mathcal{S}^-$ so that the fields are everywhere continuous.
We may thus write $\mathcal{S}=\mathcal{S}^+\cup\mathcal{S}^\delta\cup\mathcal{S}^-$ and, in analogy with Eqs~\eqref{eq:S=S+S} and \eqref{eq:C=C+C}, we have:
\begin{align}
&\int_\mathcal{S}\mathbf{A}\cdot\hat{\boldsymbol{n}}dS=\int_{\mathcal{S}^+}\mathbf{A}^+\cdot\hat{\boldsymbol{n}}dS+\int_{\mathcal{S}^-}\mathbf{A}^+\cdot\hat{\boldsymbol{n}}dS+\int_{\mathcal{S}^\delta}\mathbf{A}^\delta\cdot\hat{\boldsymbol{n}}dS~,\label{eq:S=S+S+S}\\
&\oint_\mathcal{C}\mathbf{A}\cdot\hat{\boldsymbol{t}}dl=\oint_{\mathcal{C}^+}\mathbf{A}^+\cdot\hat{\boldsymbol{t}}dl+\oint_{\mathcal{C}^-}\mathbf{A}^-\cdot\hat{\boldsymbol{t}}dl+\oint_{\mathcal{C}^\delta}\mathbf{A}^\delta\cdot\hat{\boldsymbol{t}} dl~.\label{eq:C=C+C+C}
\end{align}
Note that there is no boundary term analogous to the last one of Eq.~\eqref{eq:C=C+C} in Eq.~\eqref{eq:C=C+C+C} as we have assumed that $\mathbf{A}$ is everywhere continuous. Using the above into Faraday's law Eq.~\eqref{eq:Faraday}, we may use Reynolds's theorem in all three regions in combination with Eq.~\eqref{eq:S=S+S+S} to arrive to:
\begin{equation}
-\int_{\mathcal{S}(t)}\frac{\partial \mathbf{B}}{\partial t}\cdot\hat{\boldsymbol{n}}dS=\oint_{\mathcal{C}^+(t)}\mathbf{E}^+\cdot\hat{\boldsymbol{t}}dl+\oint_{\mathcal{C}^-(t)}\mathbf{E}^-\cdot\hat{\boldsymbol{t}}dl+\oint_{\mathcal{C}^\delta(t)}\mathbf{E}^\delta\cdot\hat{\boldsymbol{t}}dl~.
\label{eq:Faraday3regions}
\end{equation}
If, at this point, we assume that the intermediate transition region is very thin, corresponding to the limit $\delta\rightarrow0$, the last term of Eq.~\eqref{eq:Faraday3regions} goes to zero. Then, in order to combine the two remaining contour integrals on the right-hand side, we must use the formula of Eq.~\eqref{eq:C=C+C} to account for a possible discontinuity of $\mathbf{E}$ at the junction between the two regions which, in the limit $\delta\rightarrow0$, is located at $\sigma$. Using Stokes's theorem, we arrive at:
\begin{equation}
\int_{\mathcal{S}(t)}\left(\frac{\partial \mathbf{B}}{\partial t}+\mathbf{\nabla}\times\mathbf{E}\right)\cdot\hat{\boldsymbol{n}}dS=\int_\sigma[\mathbf{E}]^+_-\cdot\hat{\mathbf{t}} dl~,
\label{eq:Faradaysigmashear}
\end{equation}
where we have again replaced $\hat{\boldsymbol{t}}$ by its constant value $\hat{\mathbf{t}}$. By analogy with Eq.~\eqref{eq:Faradaysigma}, we see that Eq.~\eqref{eq:Faradaysigmashear} implies that:
\begin{equation}
[\mathbf{E}]^+_-\cdot\hat{\mathbf{t}}=0~,
\label{eq:EJCnshear}
\end{equation}
which is equivalent to (C.I).

\end{appendix}

\bibliography{bibliography}

\end{document}